\documentclass[twocolumn,showpacs,preprintnumbers,amsart,amsmath,amssymb,preprintnumbers,prl]{revtex4}

\usepackage{graphicx}
\DeclareGraphicsExtensions{.jpg,.pdf}
\usepackage{dcolumn}
\usepackage{bm}

\begin{document}

\title{Numerical Test of Born-Oppenheimer Approximation in Chaotic Systems}

\author{Jeong-Bo Shim$^{1}$}
\author{Mahir S. Hussein$^{1,2}$}
\author{Martina Hentschel$^{1}$}

\affiliation{$^{1}$Max-Planck Institute for the Physics of Complex
  Systems, N\"othnitzer Str. 38, Dresden, Germany\\
$^{2}$ Instituto de F\'{\i}sica, Universidade de S\~{a}o Paulo, C.P. 66318,
S\~{a}o Paulo, SP, Brazil}

\date{\today}

\begin{abstract}
We study the validity of the Born-Oppenheimer approximation in chaotic
dynamics. Using numerical solutions of autonomous Fermi accelerators,
we show that the general adiabatic conditions can be interpreted as
the narrowness of the chaotic region in phase space.   
\end{abstract}

\maketitle
\section{I. Introduction}
The Born-Oppenheimer approximation is a method which
deals with a coupled system of heavy and light objects\cite{born}. From atomic physics to nuclear physics, 
it has played an important role. Moreover, the importance of this method gets greater as the 
necessity of the adiabatic control of quantum states arises in the quantum information science\cite{qc}. 

Nevertheless, the criteria of the validity of the method is not well
defined yet, even still controversial in some aspects\cite{marzlin,
  tong}. Therefore, more theoretical work is needed through
applications to specific models.  

In this paper, we focused on the effect of chaotic dynamics in the Born-Oppenheimer approximation
method. Since the Born-Oppenheimer approximation basically deals with
coupled systems, there is always the
 possibility of non-integrability generating chaotic dynamics\cite{lichtenberg}. For this study, we choose
the Fermi accelerator as a model of Hamiltonian chaos\cite{fermi}.

Originally, the Fermi accelerator has been proposed to explain the origin of fast cosmic rays
 and developed further by Ulam\cite{ulam}. However, this model has had more importance as a standard
 Hamiltonian system to exhibit chaotic dynamics. Because of its simple
 composition and yet rich
dynamics, it has been intensely studied both classically\cite{lichtenberg, schmelcher} and quantum mechanically\cite{seba, bluemel, schleich}. The feature of harmonically driven barriers is frequently encountered in 
various physical situations\cite{schmelcher2, girvin}, hence it provides very useful physical insights in spite of its simple structure. Furthermore, physical features of the model have been realized experimentally in an atomic system\cite{exp}.  

 Besides the considerations of various regimes, several varieties
  of the Fermi accelerator have also been proposed, including
  relativistic models, of the type which embodies thermodynamic
  considerations\cite{pustyl} and mechanical Fermi accelerator with damping\cite{luo}. Our model can however be contrasted with the  previous ones in terms of energy conservation. While the general Fermi accelerator has a periodic external driving force, the two moving objects in our system simply exchange energy with each other, but always maintaining the total energy conserved.( In the relativistic regime,
 the system can have a constant energy effectively in spite of the driving, but it is conditionally given. See \cite{pustyl}) Accordingly, this model may be called an `Autonomous Fermi 
Accelerator'(AFA), and the property of energy conservation enables us to calculate the exact
 eigenstates. With this set of eigenstates, we analyze the dynamics of the AFA and derive the condition for the validity of the Born-Oppenheimer
 approximation in the chaotic regime.

\begin{figure}
\centering
\includegraphics[width=7cm]{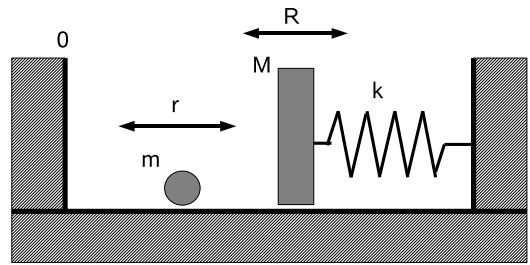}
\caption{Schemetical diagram of the autonomous Fermi accelerator}\label{fig1}
\end{figure}

This paper is organized as following. In Sec.~II, we describe the autonomous Fermi
 accelerator schematically and present the analysis of its classical dynamics  by phase space
 portraits. The phase space of the system shows under what conditions the chaotic dynamics
 sets in. Then we apply the Born-Oppenheimer approximation to the system in Sec.~III.
 We then compare the approximated calculation with the exact numerical solution which is
 obtained in Sec.~III. Through this comparison, the connection between the conventional 
criteria of the Born-Oppenheimer approximation and the chaotic dynamics can be derived.

\section{II. Classical Dynamics}\label{sec2}

\begin{figure}
\centering
\includegraphics[width=9cm]{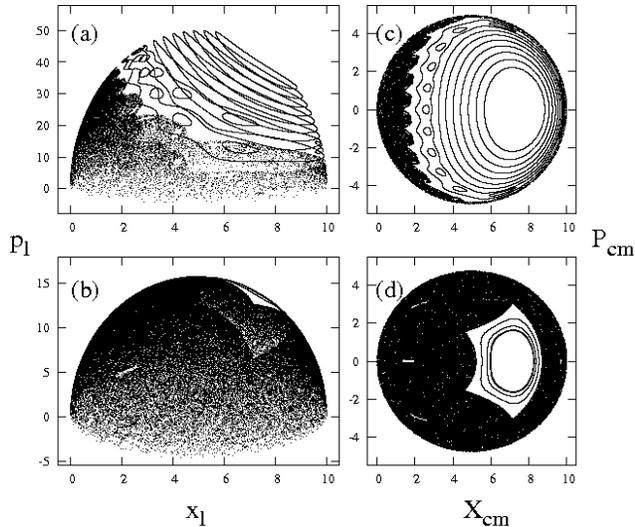}
\caption{Phase space portraits of the autonomous Fermi accelerator. The plots are constructed by recording dynamical phases at each collision between the light particle and the heavy wall : (a), (b) Plots of dynamical phases of a light particle. (c), (d) Same plots of a center of masses. The mass ratio($\eta$) and the total energy($E$) for each plot are assigned as $(\eta = 0.01$, $E = 12.0)$ for (a) and (c), and $(\eta = 0.1$, $E = 12.0)$ for (b) and (d).}\label{fig2}
\end{figure}

The autonomous Fermi accelerator consists of a free particle confined between two infinite
 potential barriers. Both barriers are not penetrable and one of them has a finite mass
 and moves in a harmonic potential, while the other one is fixed. The schematical description
is shown in Fig.~\ref{fig1}.

The Hamiltonian of the autonomous Fermi accelerator is given by 
 
\begin{eqnarray}
H(P, p, R, r) = \frac{P^2}{2M} + \frac{p^2}{2m} + \frac{1}{2} k ^2 (R - R_0)^2~,
\end{eqnarray}

where $M$, $P$ and $R$ are a mass, momentum and postion variable for the oscillating heavy
 wall, while $m$, $p$ and $r$ are those for the free particle. We set the fixed wall at $R=0$
 and the minimum point of the harmonic potential at $R_0 = 5$. Also, we assume that every collision between the particle and the wall is elastic.

First, we analyze the classical dynamics of this system by Poincare surface of sections. The
 surface of section can be constructed stroboscopically by sampling the dynamical status of
 either the particle or the wall at each collision. The obtained results with different parameters
 are presented in Fig.~\ref{fig2}. The right column of the figure shows the phase space plots of the
 light particle and the left one shows those of the center of masses.

In this figure, we can notice that the feature of energy conservation is vividly shown;
 namely that all dynamical phases are restricted on a spherical surface given by
 
\begin{eqnarray}
E = \frac{P^2}{2M} + \frac{p^2}{2m} + \frac{1}{2} k ^2 (R - R_0)^2~. 
\end{eqnarray}

 As Fig.~\ref{fig2} shows, the phase space of the system is quite clearly divided 
into two parts, a stochastic and a regular part. The border between them is conditioned by
 the existence of the double collision. Here, the double collision means the situation
 that the light particle is hit by the heavy wall twice
 until it reaches the fixed wall. The presence of this type of collision diverges the light
 particle's trajectory drastically, and results in the stochastic dynamics. Apparently, 
it can be a necessary condition for the double collision that the heavy wall is faster than
the light particle. Fig.~\ref{fig2} shows such tendency well. The stochastic region is 
located around the equator of the sphere in the phase space, while the regular island is  
around the pole.

We also mention a structural detail in phase space: Around the border between 
the chaotic and the regular region there exists ``Stochastic Web Structure"\cite{zaslavsky}. However, this
structure is ignorable in the regime of wave mechanical parameters in this study. 


The structural properties of the phase space, including the portion of the regular and
 chaotic region, is controlled mainly by the mass ratio($\eta$). By the comparison of
 (a) and (b) in Fig.~\ref{fig2}, we can confirm the tendency that the regular island
 is shrinked as $\eta$ is increased. Also the phase space is affected by the total energy($E$),
 but the portion of regular island is not changed as much as by the mass ratio,
 according to our numerical study.  
 
\section{III. Born-Oppenheimer Approximation} \label{sec3}
If the system is quantized, the total Hamiltonian becomes 

\begin{eqnarray}
H = - \frac{\hbar ^2}{2M} \frac{\partial ^2}{\partial R^2} - \frac{\hbar^2}{2m} \frac{\partial ^2}{\partial r^2} + \frac{1}{2} k (R - R_0)^2  
\end{eqnarray}

and the Schr\"odinger equation and the boundary condition are given by 

\begin{eqnarray}
\label{scheq}
H \Psi(r,R) = E \Psi(r, R)~,
\end{eqnarray}
\begin{eqnarray}
\Psi (r=0, R) = \Psi(R \leq r, r) = 0 ~.
\end{eqnarray}

Application of the Born-Oppenheimer approximation means to fix all parameters which are given
 instantaneously by the heavy object. In our system, this can be done by fixing
 the distance between the  walls, to be $R$, at every moment. Then, a
 set of basis states can be constructed as follows:

\begin{eqnarray}
\Psi (R, r) = \sum_{n} \xi _{n}(R) \phi _{n} (r, R) 
\end{eqnarray}

where

\begin{eqnarray}
\label{oned}
\phi _{n} (r, R) = \sqrt{\frac{2}{R}} \sin{\frac{n \pi r}{R}}~.
\end{eqnarray}

Using the completeness of the basis set above, the Schr\"odinger equation~(\ref{scheq}) can be
 reduced in the following way:

\begin{align}
\label{boapp}
\nonumber \sum_{n^\prime} < \phi _n|- \frac{1}{2M} \frac{\partial ^2}{\partial R ^2} |\phi _{n^\prime}> \xi _{n^\prime} (R)\\ + \frac{1}{2} k (R-R_0)^2 \xi_{n} (R) + \epsilon _{n} \xi (R) = E \xi_{n} (R)~,
\end{align}

where $\epsilon = (\hbar n \pi)^2 / 2m R^2$. 

If Eq.~(\ref{oned}) is inserted into (\ref{boapp}), one is able to expand the equation with
 the mass-ratio, $\eta(=m/M)$ as a parameter. By taking up to the first order of this expansion,
 the following equation is obtained,

\begin{align}
\label{approximation}
\nonumber ( - \eta \frac{\partial ^2}{\partial R ^2} + \eta \frac{M k}{\hbar^2}(R - R_0)^2 - \frac{n^2 \pi^2}{R^2} ) \xi_{n} (R) \\ = \eta \frac{2ME }{\hbar^2} \xi_{n}(R) ~.
\end{align}

 Consequently, the governing equation is reduced to a simple
1-dimensional equation of motion with a stationary bounding potential and the third term of
 the equation can be thought of as a pressure generated by the frequent collisions of the light particle.

According to the theory of Hamiltonian chaos, there is no possibility of chaos if
 we consider the dynamics at the pure classical level. However, if we take the level transition into
 account, the effective potential can be thought of to be time-dependent, and it
seems more reasonable as the counterpart of the classical chaotic system. Along this line of
 reasoning, if an eigenmode is corresponding to chaotic dynamics, we can make an assumption that
the lowest order of approximation would fail to describe the
eigenmode. The validity of this assumption,
 will be discussed in the next section through a discussion of the exact solution which is
 numerically obtained. 

\section{IV. Numerical Study}\label{sec4}
\begin{figure}
\includegraphics[width=10cm]{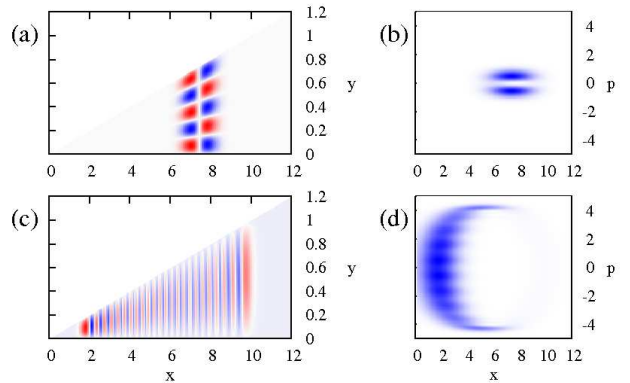}
\caption{Configurational Plots of Eigenmodes at $\eta = 0.01$. : (a) Regular mode : $E=12.38$ (c) Chaotic mode : $E=12.29$. (b), (d) : Corresponding Husimi distributions to (a) and (c), respectively} \label{fig3}
\end{figure}

As mentioned in the introduction, the exact solutions of the system are also available.
To calculate them, the following transform is performed.  

\begin{eqnarray}
R = x~, ~~~ r = \eta~y~.
\end{eqnarray}

Then the Schr\"odinger equation becomes

\begin{eqnarray}
(- \frac{\hbar ^2}{2M} \nabla ^2 + \frac{1}{2} k (x-x_0)^2) \Psi (x,y) = E \Psi (x,y)~.
\end{eqnarray}

As the result of the transformation, the 1-dimensional dynamics of the system becomes a 2-dimensional dynamics in a triangular well, the three sides of which are two infinite potential barriers along the x-axis and $y = \tan \theta x$, and one harmonic potential barrier to the x-direction.
 Here, the angle $\theta$ between the two infinite barriers is given by 

\begin{figure}
\includegraphics[width=10cm]{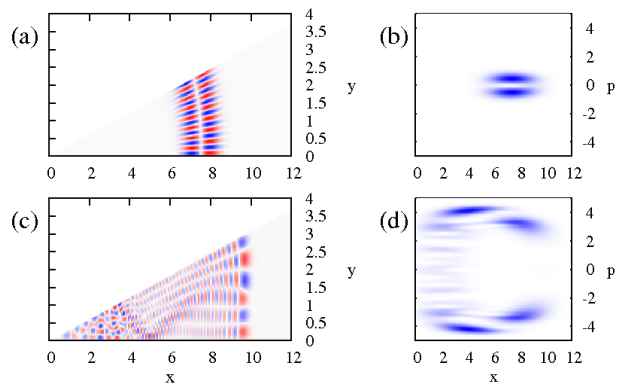}
\caption{Configurational Plots of Eigenmodes at $\eta = 0.1$: (a) Regular mode : $E=12.22$ (c) Chaotic mode : $E=12.19$. (b), (d) : Corresponding Husimi distributions to (a) and (c), respectively.} \label{fig4}
\end{figure}

\begin{eqnarray}
\theta = \tan ^{-1} \sqrt{\eta}~,
\end{eqnarray}

where $\eta$ is the mass ratio($=m/M$).  It should be noted that
  $\theta$ would reach $\pi /2$ as $\eta$ approaches infinity. In this
  work, we are concerned only with the normal situation of Born-Oppenheimer approximation, namely that $M$ is much larger than $m$.

Then the boundary condition is given as

\begin{align}
\label{transeq}
\nonumber \Psi(x, y=0)=0~,\\ 
\Psi(x, y=\tan \theta x) =0~,\\
\nonumber \Psi(x \rightarrow \infty, y) =0~.
\end{align}

Using a numerical approach, we obtain all possible eigenvalues and corresponding eigenmodes
 with two different mass-ratio parameters, $\eta = 0.01$ and $0.1$.
 As the numerical algorithm, `Finite Element Method'\cite{fem} is used.

Fig.~\ref{fig3} and Fig.~\ref{fig4} show configurational plots and Husimi distributions
 of eigenmodes at $\eta = 0.01$ and $\eta = 0.1$, respectively. Using the normal derivative of 
configurational mode distributions along $ y = \eta x$, we can calculate the Husimi distributions,
 and these are corresponding to the Poincar\`e surface of section of a center of mass.
 Hence, the corresponding dynamics for each mode can be confirmed by
 comparison of the Husimi distributions with the classical phase spaces in Fig.~\ref{fig2}.
In Fig.~\ref{fig3} and \ref{fig4}, (a) and (b) of each figure shows a mode which is
 localized the regular regime in the phase space, and (c) and (d) shows ones in
 the chaotic regime.

From the obtained computational results, the validity of the Born-Oppenheimer approximation is 
investigated. For this purpose, the Fourier analysis is performed to find the eigenmode composition
 of the light particle.

\begin{eqnarray}
F(n) = \int^{\infty}_{0} \int^{\eta x}_{0} \Psi(x,y) \sin (n \pi y/x) dy dx
\end{eqnarray}

\begin{figure}
\includegraphics[width=8cm]{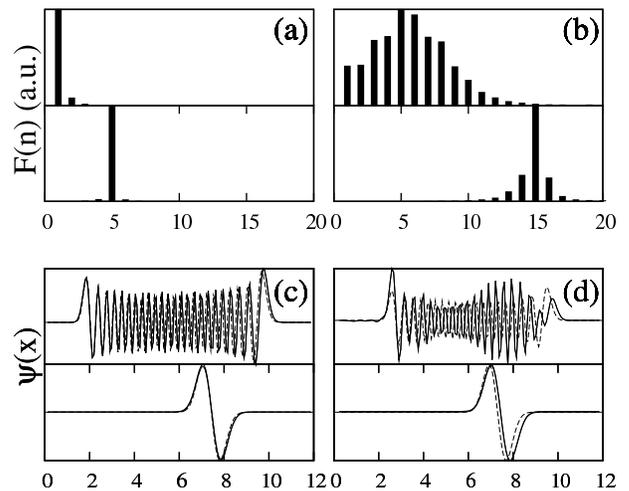}
\caption{(a) The analysis of the modes in Fig.~\ref{fig3} ($\eta = 0.01$) under the instantaneous basis of the light particle. (b) The same one for Fig.~\ref{fig4}. In each figure, the upper distribution is for modes in the chaotic region and the lower one is for ones in the regular region. (c), (d) The extracted wave functions of the moving wall(solid line) and the approximated wave functions(dashed line)
 } \label{fig5}
\end{figure}

The result of the analysis shows the different features of distributions,
 depending on the underlying dynamics. For $\eta = 0.01$, the distribution is well localized
regardless to the dynamics (see Fig.~\ref{fig5}(a).). For $\eta = 0.1$, however, the distribution
 shows the interesting features depending on the
 dynamics (Fig.~\ref{fig5}(b)). The mode which is placed
 in the chaotic region shows a broad distribution, with tails reaching the regular region, but
 the one in the regular region still has a well localized distribution. 

Considering the scheme of the Born-Oppenheimer approximation, modes with the high localization are
 expected to be well-approximated. To confirm this, we extract the wave
function of the moving wall with the highest level ($\bar{n}$) in each distribution in the following way,

\begin{eqnarray}
\psi(x) = \int^{\infty}_{0} \Psi(x,y) \sin(\bar{n} \pi y /x)) dy ~.
\end{eqnarray}

Fig.~\ref{fig5} presents a comparison between the wave function extracted from the exact
 wave function in Fig.~\ref{fig3}(c) and the one approximated by the Eq. (\ref{approximation}).
At $\eta = 0.01$, the extracted wave function agrees very well with the approximation,
 regardless to where in phase space region a mode is situated (Fig.~\ref{fig5}(c)). In contrast,
 the mode in the chaotic region for $\eta =0.1$ does not show a good agreement
 with the approximation, while the one in the regular region still abides by the aforementioned
approximation (Fig.~\ref{fig5}(d)). Of course, the disagreement is due to the level transitions during the dynamics. 

Now, it is necessary to consider the general criteria for the variation of Born-Oppenheimer
 approximation\cite{ali,mackenzie,adiabatic}, which is given as follows,

\begin{eqnarray}
\label{ineq}
 \hbar |<n|\dot{m}>|  \ll |E_n - E_m|~.
\end{eqnarray}

This inequality simply guarantees that the heavy object is moving slowly enough to prevent
 level transitions. If we apply this inequality to our system, the condition of this inequality 
is not changed much by the dynamical status. In other words, the ratio between the left term
 and the right term in Eq.(\ref{ineq}) is given as around  $0.01$ for $\eta = 0.01$ and
 around $0.2$ for $\eta = 0.1$, no matter whether a mode is chaotic or regular.

Considering the near-integrability in the case of regular dynamics, it is obvious that the simple
 approximation works well in the regular region, because we can define good quantum
 numbers for each degree of freedom.

In addition, we can find the dynamical interpretation of the inequality (\ref{ineq}). In 
our system, the satisfaction of (\ref{ineq}) guarantees that the chaotic region is small enough
in comparison to Planck's constant $\hbar$. Therefore, it is impossible that more than two 
modes are situated in the chaotic region. Accordingly, no transition is possible. 

\section{V. Conclusion}
In this work, we study the validity of the Born-Oppenheimer approximation in the presence
 of chaotic dynamics. Especially, the criteria under which the approximation is working is interpreted
 from the dynamical point of view. We apply the approximation method to the autonomous
 Fermi accelerator which has 
a clear division of regular dynamics and chaotic dynamics in its phase space. In the regular
region, the approximation is obviously well suited to calculate the eigenmodes, whereas
it is just conditionally suited in the chaotic region. However, when the dynamics of the system satisfies the general condition for the approximation,
 that is, the evolution of the system is slow enough to prevent the level transitions,
 the area of the chaotic region becomes small enough in comparison to Planck's constant.
 Thus, the chaotic dynamics can not be resolved by wave functions. Thereby, the Born-Oppenheimer
approximation can approximate the wave functions very well in spite of
the chaotic nature of the dynamics. 

It is important to remark that in the region where the Born-Oppenheimer
approximation is working, we can apply another approximation to our system.
Then we can follow the procedure of Ref.~\cite{hussein} to construct the density matrix
and the reduced density matrix of the interesting subsystem (light particle or
oscillating wall). We leave this for a future work. We also remark that it would
be quite instructive to study higher order corrections to the Born-Oppenheimer
approximation \cite{ali,mackenzie} in connection with chaotic dynamics. We also leave this
for a future study.
\\\\
This work was supported in part by the Max-Planck-Institute for the
Physics of Complex Systems (MPIPKS) in Dresden, the Emmy-Noether-Programme of the DFG (Germany) and by the Brazilian agencies, CNPq and FAPESP.

\end{document}